\documentclass[prb,apsrev,floatfix,twocolumn]{revtex4-1}

\usepackage{amsmath}
\usepackage{graphicx,color}
\usepackage{amsfonts}
\usepackage{amssymb}
\usepackage{graphicx}%
\providecommand{\U}[1]{\protect\rule{.1in}{.1in}}

\begin{document}

\title{Electron dynamics method using a locally projected group diabatic Fock matrix 
for molecules and aggregates}

\date{\today}
\author{Takehiro Yonehara}
\email{takehiro.yonehara@riken.jp}
\author{Takahito Nakajima}
\email{nakajima@riken.jp}
\affiliation{RIKEN Center for Computational Science, Kobe 650-0047, Japan}

\begin{abstract}
We propose a method using reduced size of Hilbert space to describe 
an electron dynamics in molecule and aggregate based on our previous theoretical scheme  
[ T. Yonehara and T. Nakajima, J. Chem. Phys. \textbf{147}, 074110 (2017) ].
The real-time time-dependent density functional theory 
is combined with newly introduced projected group diabatic Fock matrix. 
First, this projection method is applied to a test donor--acceptor dimer, namely, 
a naphthalene--tetracyanoethylene with and without initial local excitations and light fields.  
Secondly, we calculate an absorption spectrum of five-unit-polythiophene monomer. 
The importance of feedback of instantaneous density to Fock matrix is also clarified. 
In all cases, half of the orbitals were safely reduced 
without loss of accuracy in descriptions of properties. 
The present scheme provides one possible way to investigate and 
analyze a complex excited electron dynamics in molecular aggregates 
within a moderate computational cost.

\end{abstract}

\maketitle




\section{Introduction}

Recently, the authors proposed a concise method for describing 
the quantum dynamics of excited electrons traveling over 
constituent molecules in a molecular aggregate system 
by utilizing a group diabatic Fock (GDF) matrix \cite{gdf-eld}. 
The construction scheme for this GDF matrix using atom centered Gaussian basis set
was practically demonstrated by Thoss {\it et al.} \cite{Kondov-Thoss-POD} 
based on the idea of diabatization of block diagonalization of Hamitonian 
that was originally introduced by K\"{o}ppel et al.\cite{Diab-block-diag} 
Before that, the applications of this block diabatization concept 
for a practical calculation can be found in the works on electron conduction problems. 
\cite{elec-conduc-Block-Diab-Domcke,elec-conduc-Block-Diab-Ratner}
%
%
%
%
%
%

This block diabatic scheme for the electron dynamics 
provides an intuitive understanding of charge and exciton migration as 
the quantum mechanical transport of electronic energies
in a molecular assembly subject to the inherent electronic propensities 
of the site molecules, starting from any prepared type of initial local excitations 
considering the light-electron couplings.

In the present article, on the basis of the characteristics 
of a nearly block structured diabatic representation in the GDF method, 
we introduce a compact representation within an extracted Hilbert subspace.  
The aim of this paper is two-fold:  
(1) to introduce a scheme for reducing the computational cost
    in keeping with an accurate description of the dynamics, 
and  
(2) to investigate how many local orbitals are required for a description of dynamics     
    and absorption spectrum toward an interpretation of electron dynamics. 

Note that the group diabatic representation treated here 
does not lead by itself to a reduction of computational cost 
by describing the dynamics with a small subset of the full orbital space.
In fact, the GDF matrix has the same dimension as the original Fock matrix before 
the block diagonalization.  
(1) and (2) are obtained by using a projection scheme introduced here 
as an extraction of local group diabatic orbitals 
associated with a size of the Hilbert space needed for 
a description of electron dynamics. 
The reduction of calculation cost is associated with the reduction of
the number of extracted local group orbitals 
as a basis set for constructing matrix representations of electronic operators 
utilized in a calculation of electron dynamics. 

%
 This article improves upon our previous work by providing and 
assessing a procedure for reducing the computational cost of electron dynamics calculation 
using the GDF method, which is combined with 
the real-time time-dependent density functional theory (RT-TDDFT)
\cite{
book-lec-note-TDDFT-Gross,book-fundamental-TDDFT-Gross,book-TDDFT-Ullrich,rev-RT-TD-elec-X-Li}
under the adiabatic approximation for electron functionals. 
%

The point is that electron migration of chemical interest  
over a molecular aggregate under moderate light conditions 
occurs in a sub Hilbert space consisting of a low number of excited states 
described by molecular orbitals within a relevant but not very large energy range 
around the Fermi energy of a system.

It is instructive to compare other studies with our present method. 
There are many studies intensively investigating molecular interactions and their effects 
on excited energy transfer proceeding in excited molecular aggregates.
%
%
%
%
%
%
%
\cite{Thoss-TiO2,Lan,Gao,Shi,Herbert-FrenkelDavydov,fujimoto-TDFI,fujimoto-yang-DFI,CR2017-Charge-Transport,Akimov-Prezhdo-JACS2014-NonAdiabatic-CT-SF-pentacene-c60,Wang-PCCP2015-MQC-for-CT,Spencer-Blumberger-JCP2016-MQC-for-CT,Spencer-Blumberger-JCP2016-FOB-SH,Elstner-JCTC2016-temp-dep-CT}
For example, Futera {\it et al.}\cite{Futera-elec-coupling-POD} successfully utilized and assessed 
the GDF method originally named a projection operator diabatization scheme\cite{Kondov-Thoss-POD}
to evaluate electronic coupling matrix elements with high level ab initio calculations 
in the electron-transfer process of a molecular-metal/semiconducter interface.  
However, a study on a systematic variable description 
of the excited electron dynamics in a bottom-up approach is rare.  
Compared to previous studies on excited electron transfer in molecular aggregates, 
the scheme introduced in the present article 
has the advantages of a compact description of excited electron migration 
with ab initio electronic structure calculations.
In addition our new scheme allows 
a systematic improvement of the results by enlarging the projection space.  
The most prominent feature is that our scheme is intended for a real-time dynamics of 
excited electrons in molecular aggregates in an external field 
starting from any pattern of initial local excitations prepared as 
a perturbation for the electronic state. 

In this article, we detail the procedures for constructing the projected local orbital space 
within the group diabatic (GD) representation, and then 
demonstrate numerical applications.
We examine the size of a local orbital space related to dynamics 
in a systematic way by increasing an energy range 
for projecting a diabatic local orbital space.
A naphthalene(NPTL)$-$tetracyanoethylene(TCNE) dimer is treated as a test donor$-$acceptor system. 
Additionally, we also demonstrate a convergence of absorption spectrum 
with respect to a size of orbital space using a five-unit polythiophene molecule(5UT) 
as a typical electron donor species in solar cell materials. 
%
%
To ensure the energy balance between local projected orbitals, 
we employ an energy width parameter for extracting a relevant subspace 
with the mean value of the highest occupied molecular orbital (HOMO) 
and lowest occupied molecular orbital (LUMO) energies 
in the whole system as a reference energy,   
and we do not use a scheme which requires direct orbital selection.  


In Sect. II, we explain the theoretical method for describing the electron dynamics 
based on the GDF representation within projected local diabatic orbitals,   
which is followed by numerical examples in the section III. 

%

%
%

\section{Theoretical method}

In this section, after a brief summary of the electron dynamics method \cite{gdf-eld}  
using a GD representation,\cite{gdf-eld,Lan,Gao,Shi,Thoss-TiO2,CR2017-Charge-Transport}  
we introduce a scheme for constructing a concise matrix form 
with use of projected diabatic local orbitals 
having a dominant contribution to the underlying dynamics.  
The determination of diabatic local projection orbitals playing a main role 
in the present work requires only the parameters of energy ranges for monomers 
covering the important orbitals around HOMO and LUMO 
playing in an excited electron dynamics.

\subsection{Overview of the locally projected-space group diabatic representation}

The transformation from an atomic orbital (AO) representation of physical operators 
to a GD form consists of the following two transformations being constructed sequentially: 
\begin{itemize}
\item[(i)]
     a transformation to a representation using 
     the L\"{o}wdin orthogonalized atomic basis functions and  
\item[(ii)]
     a unitary transformation made from the local orbital sets obtained by 
     the diagonalizations of block sub-matrices corresponding to 
     the predetermined monomer groups in the Fock matrix prepared in step (i).  
\end{itemize}

The overview of the process of extracting the diabatic local projection orbitals 
as the primary topic of the present article is as follows: 
\begin{itemize}
\item[(a)]
     calculate the mean of the HOMO and LUMO orbital energies of the whole system;
\item[(b)]
     set an energy range covering the local orbitals for each monomer 
     in which the value obtained in (a) is placed at the middle; 
\item[(c)]
     obtain the diabatic local projection orbitals of the monomers 
     in the energy range prepared in (b). 
\end{itemize}
We refer to the matrix representation within these projection orbitals as 
the locally projected-space group diabatic Fock (LP-GDF) representation, 
of which the details are explained in later subsections. 
     
The multiplications of matrices associated with the electronic properties 
and the analyses of the time-dependent electron density 
using the newly introduced LP-GDF representation
are carried out within the projected orbital space. 
The information related to the size of the orbital space required for 
a description of the dynamics without loss of accuracy 
provides us with insight into the sub-Hilbert space relevant to it. 

In the following subsection, for a self-contained form of the present article, 
we first summarize the electron dynamics scheme using a GD representation\cite{gdf-eld}  
and then proceed to describe how to obtain the diabatic local projection orbitals 
and how to construct compact matrix representations 
by using them as a subset of the basis functions.

\subsection{Group diabatic representation}

\subsubsection{Fock matrix in the L\"{o}wdin representation}

The first step is to prepare a Fock matrix represented 
by the L\"{o}wdin orthogonalized atomic basis function \cite{Szabo}
\begin{align} 
\widetilde{F}_{mn}    \equiv
  \langle \widetilde{\chi}_m | \widehat{F} | \widetilde{\chi}_n \rangle, 
\end{align} 
where the orthogonalized L\"{o}wdin atomic orbitals(AOs) are expressed by 
\begin{align} 
  | \widetilde{\chi}_n \rangle
 = \sum_j^{\textrm{AO}} | \chi_j \rangle (\underline{\underline{S}}^{-1/2})_{jn}  
\end{align} 
with $S_{jn} = \langle \chi_j | \chi_n \rangle$ 
being the AO overlap matrix element.  
Here $\{ \chi_n \} $ is the original basis set consisting of AOs.

\subsubsection{Localized orbitals of a subgroup}

After the classification of $\{  \widetilde{\chi}_n  \}$ into subgroups, 
e.g., monomers,  
the block structure of the Fock matrix within the L\"{o}wdin basis set 
is determined with its diagonal blocks 
$\{\underline{\underline{\widetilde{F}}}_{G_i G_i}\}$ and 
off-diagonal ones $\{\underline{\underline{\widetilde{F}}}_{G_i G_j}\}_{i\neq j}$ 
with i and j ranging from 1 to $N_g$.
$G_i$ denotes the i-th subgroup. $N_g$ is the number of subgroups in the system.  
Note that we can employ arbitrary divisions of the component atoms 
in the whole system.

The diagonalization of diagonal blocks corresponding to subgroup $G$, 
\begin{align} 
 \underline{\underline{\widetilde{F}}}_{GG}  
=
 \underline{\underline{D}}_{G}  
\,
 \underline{\underline{\overline{F}}}_{GG}  
\,
 \underline{\underline{D}}_{G}^{\dagger}, 
\end{align} 
gives rise to the unitary transformation matrix 
$\underline{\underline{D}}_{G}$, 
whose column vectors are the linear coefficient vectors of 
the localized eigenstates expanded in terms of 
the L\"{o}wdin orthogonalized atomic basis functions for the group $G$. 
The dagger symbol attached to a matrix indicates its adjoint form. 
$ \underline{\underline{\overline{F}}}_{GG} $ is the diagonal matrix 
having the eigen energies $\{\epsilon_{j,{G}}\}_{j = 1 \sim M_{G} }$
of the corresponding subgroup $G$ as its elements, and $M_{G}$ 
is the number of local basis functions spanned at group $G$.    
Here, $ {G} \in \{ {G}_1,...,{G}_{N_g} \} $.

The elements in off-diagonal blocks of the Fock matrix represented by these localized orbitals,  
associated with different groups, can take non-zero values,  
which provide a diabatic character in the representation 
with the use of the collection of these orbital sets. 
These group localized orbital sets provide a transformation matrix 
from a L\"{o}wdin representation to the GD one to be explained later.

Here, we provide a short comment on the ambiguity of group division 
in a L\"{o}wdin representation of Fock operator. 
In case using diffuse AO orbitals having far larger distribution 
than a distance between monomers, 
a center of L\"{o}wdin orbital created 
by the mixture of original AOs can become close to 
a nuclei belonging to a different group. 
Though that is one of the possible problem in the 
group assignment of L\"{o}wdin orbitals associated with 
a natural group sectoring, it does not cause any problem  
in case with distances of monomers sufficiently 
larger than AOs and also L\"{o}wdin orbitals.  
\cite{CR2017-Charge-Transport}
In this article, we do not use diffuse orbitals and 
treat a case with distance of monomers being larger than 
a spatial range in orbital distribution.

\subsubsection{Group diabatic Fock matrix}

The GD representation of the Fock operator $\underline{\underline{\overline{F}}}$, 
as one of the main ingredients in the GDF electron dynamics scheme is constructed via   
the transformation of the L\"{o}wdin representation matrix of the Fock operator
$ \underline{\underline{\widetilde{F}}} $. 
By using the already obtained unitary matrices with 
the dimensions of the local basis functions associated with groups, 
$\{\underline{\underline{D}}_{G_i}\}_{i=1 \sim N_g}$, 
this transformation is expressed by 
\cite{gdf-eld,Lan,Gao,Thoss-TiO2} 
\begin{align} 
 \underline{\underline{\overline{F}}}_{G_i\,G_j}  
=
 \underline{\underline{D}}_{G_i}^{\dagger}  
\,
 \underline{\underline{\widetilde{F}}}_{G_i\,G_j}  
\,
 \underline{\underline{D}}_{G_j},   
\label{SB-GDFmat}
\end{align} 
where i and j range from 1 to the number of groups $N_g$. 
The assembly of these sub-matrices
$ \left\{ \underline{\underline{\overline{F}}}_{G_i\,G_j} \right\}_{i,j=1\sim N_g} $
constructs the GD representation which is called GDF matrix and expressed by 
$  \underline{\underline{F}}^{\textrm{GD}}  $. 
The physical meaning of the components in this final form is as follows.
This sub-matrices
$ \left\{ \underline{\underline{\overline{F}}}_{G_i G_i}  \right\}_{i=1 \sim N_g}$ 
in the diagonal blocks correspond to the local group eigen energies, 
while 
$ \left\{ \underline{\underline{\overline{F}}}_{G_i G_j}  \right\}_{i \ne j}$,
placed at the off-diagonal blocks, 
describes the interactions between different groups.
Note that, in this transformation,  
the information included in the AO, L\"{o}wdin, and GDF representations are 
the same and no approximation is applied.

\subsubsection{Transformation from the AO representation to the GD representation}

A matrix representation of any observable operator $\hat{O}$
in terms of the constructed GD basis set, 
$\underline{\underline{O}}^{\mathrm{GD}}$, 
is related to that of the original AO basis set, 
$\underline{\underline{O}}^{\textrm{AO}}$, 
as 
\begin{align} 
\underline{\underline{O}}^{\mathrm{GD}} 
=
\underline{\underline{U}}^{\dagger} \,
\underline{\underline{O}}^{\textrm{AO}} \,
\underline{\underline{U}}, 
\label{trans-obs}
\end{align} 
where 
$
\underline{\underline{U}} \equiv 
\underline{\underline{S}}^{-1/2} \underline{\underline{W}} 
$.
Here, the diagonal block in the sub-transformation matrix $ \underline{\underline{W}}$
is given by 
$
\underline{\underline{W}}_{G_i G_i} \equiv \underline{\underline{D}}_{G_i} 
$
for $i$,  
while 
the off-diagonal one is defined as 
$
\underline{\underline{W}}_{G_i G_j} \equiv \underline{\underline{0}} 
$.
%
The Fock matrix obeys the same transformation rule 
and is obtained by setting $\hat{O}=\hat{F}$ in the above equations, 
where we know that 
$ \underline{\underline{F}}^{\textrm{AO}} = \underline{\underline{F}} $ and 
$ \underline{\underline{F}}^{\mathrm{GD}} = \underline{\underline{\overline{F}}} $. 

On the other hand, the transformation of the density matrix 
from the original AO basis set to the GD one is written as 
\begin{align} 
\underline{\underline{\rho}}^\mathrm{GD} 
=
\underline{\underline{U}}           \,
\underline{\underline{\rho}}^\mathrm{AO}   \,
\underline{\underline{U}}^{\dagger}. 
\label{trans-dens}
\end{align} 
Note that the unitarity of $\underline{\underline{W}}$ assures 
total electron conservation with respect to this transformation, 
\begin{align} 
\textrm{Tr} 
\left[ \underline{\underline{\rho}}^{\mathrm{AO}} \underline{\underline{S}} \right]
= \mathrm{Tr} \left[ \underline{\underline{\rho}}^\mathrm{GD} \right]. 
\end{align}

\subsubsection{State coupling}

We can obtain the essential elements needed for the construction of light--electron coupling 
by setting $ \hat{O} = \hat{\bf r} $,  $ \partial_{\bf r} $ in the previous subsection.
Here, boldface denotes a vector in a three-dimensional Cartesian space, 
and ${\bf r}$ denotes a composite variable of the electron position in 
three-dimensional space.
The first and second operators are responsible for the light--electron coupling 
in length and velocity forms. \cite{Faisal-LG-VG,ACP-laser}
Here, we neglect the nonadiabatic coupling and 
molecular motion to allow us to focus on an examination of 
the electron dynamics scheme using projected local diabatic group orbitals. 

%
%
Light-electron couplings within GD representation 
are expressed by 
$
\underline{\underline{L}}^\mathrm{GD}  
=
+ e \underline{\underline{{\bf r}}}^\mathrm{GD} {\bf E} 
$
for the length gauge and 
$
\underline{\underline{L}}^\mathrm{GD}  
=
- i \hbar \dfrac{e}{c} {\bf A}
\underline{\underline{\partial_{\bf r}}}^\mathrm{GD} 
$
for the velocity gauge.
\cite{Faisal-LG-VG,ACP-laser}
${\bf E}$ and ${\bf A}$ denote respectively 
the three-dimensional electric field vector and 
electromagnetic field vector potential, 
which generally depend on a point in a three-dimensional space.  
$\underline{\underline{{\bf r}}}^{\textrm GD}$ and 
$\underline{\underline{\partial_{\bf r}}}^{\textrm GD}$
denote the GD presentation matrices respectively 
for $\hat{\bf O}={\bf r}$ and $\partial_{\bf r}$ in Eq. (\ref{trans-obs}).
%
In this study, we employ the length gauge. 
The details on the treatment for these coupling matrices 
with further approximations applied in dynamics calculation 
are given in Subsect. \ref{section-time-propagation}.  


\subsection{Locally projected space made from subsets of local site orbitals
\label{subsec-proj}}

This subsection is an essential part of the present article proposing the LP-GDF scheme. 
Here, we provide the concrete procedures used in it 
and the corresponding mathematical expressions. 

Let us consider the diagonal block of the Fock matrix within the GD representation, 
$\left\{\underline{\underline{\widetilde{F}}}_{G_i G_i}\right\}$ 
and the associated local orbital energies
$\{\epsilon_{j,G_i} \}_{j=1\sim N_{G_i}}$ 
corresponding to the i--th group site. 
In a situation where the interactions between group monomers 
are weak compared to those among the atoms in each monomer,  
we can safely employ the referential local ground state,
where all of the electrons assigned to a local site are filled 
in ascending order from the lowest-energy local orbital.

The excited electron dynamics in a molecular aggregate system under  moderate sunlight conditions 
is expected to proceed in low-excited-state manifolds 
constructed from the local molecular orbitals in a relevant but not very large energy range 
around the Fermi energy of the system. 
In a situation involving rather strong light field or molecular interaction, 
the Hilbert subspace required for a description of the dynamics will become large. 
It is worthwhile to examine the size of the subspace relevant to the excited electron dynamics 
by varying the type of molecular interactions  
such as light-matter coupling and initial local excitations. 

Thus, in order to examine the size of the orbital space needed for 
a description of the electron dynamics at a sufficient accuracy, 
we establish a procedure for projecting a Hilbert subspace around the Fermi energy 
of the whole molecular aggregate by using a width parameter for the energy range as follows: 
\begin{itemize}
\item[(i)]
      Set an energy width $\Delta \epsilon_i$ 
      covering the sub-space consisting of local orbitals orbitals 
      for the $i$-th monomer, $G_{i}$.   
      Here $i$ ranges from 1 to $N_g$.
\item[(ii)]
      Calculate the mean $\eta$ of the HOMO and LUMO energies of the whole system.
\item[(iii)]
      Extract a subset of group diabatic local orbitals in the energy range 
      $ [ \eta - \Delta\epsilon_i/2,   \eta + \Delta\epsilon_i/2 ] $ 
      from the whole orbital space.
      We refer to this as a projection of the diabatic local orbitals.
\item[(iv)]
      Construct minor matrices as representations of the electronic operators 
      within the projected orbital basis set obtained in (iii).
\item[(v)]
      Follow the same procedure as that in the GDF electron dynamics scheme 
      other than the projection introduced here. 
      This means that time propagation and property analysis are carried out 
      using the obtained small size matrices. 
\end{itemize}
The mathematical form of LP-GDF procedure is compactly given in Appendix \ref{app-LP-GDF}. 

In this article, we employ a version of projection independent on the orbitals 
during a time propagation but constructed from the initial orbitals in dynamics.
The orbital projection scheme using an energy width 
allows for the systematic and natural treatment for general situations involving the
unknown energy orders and spatial localities of the GD orbitals in molecular aggregates. 
%

Note that 
to ensure the energy balance between local projection orbitals, 
we employ an energy width parameter for extracting a relevant subspace 
with the mean value of the highest occupied molecular orbital (HOMO) 
and lowest occupied molecular orbital (LUMO) energies 
in the whole system as a reference energy,   
and we do not use a scheme which requires direct orbital selection.

In the following numerical demonstrations, 
for focusing on this essential point in the proposed scheme, 
we employ a parameter for the energy range independent of the monomers, namely,  
$\Delta \epsilon_i$ = $\Delta \epsilon$, 
which is expressed by $E_{\rm bw}$ in the figures.
The resultant total number of projection orbitals of the monomers 
is denoted by $N_{\rm proj}$ throughout the present article.


\subsection{Initial density matrix: local excitation and electron filling}
\label{init-dens}

The initial density matrices in the GDF representation are prepared so that 
the diagonal elements in each diagonal block of the corresponding monomer 
should be occupied up to the number of electrons assigned to this monomer. 
A simple example can be found in our previous article.\cite{gdf-eld}
Though an off-diagonal filling responsible for the initial coherence is also possible, 
for simplicity, we consider only the diagonal part in 
the preparation of the initial state of the density matrix.
This issue will be reported in our future article. 
In the article, we treat the case of the spin-restricted model 
within the GDF representation as explained below. 
This is not rigorously the same as that within the canonical 
(KS or HF) orbital representation. 
In a restricted case with the same spatial orbitals 
for different alpha and beta spins in this model,
the occupancy of the GDF local orbitals of each monomer is up to half of 
the number of electrons assigned to this group 
from the lowest energy orbital. 
Therefore, in a strict sense, the initial density matrix mentioned above  
differs from that of the true ground state of the whole system.  
In fact, this does not cause any problem for examining the migration dynamics 
of charge and electronic excitations over the constituent monomers in an assembly.\cite{gdf-eld}

Generally, we can make any type of excitation configuration 
starting from the reference occupations of the GD orbitals.  
If we want an initial density associated with an excess or a deficiency of electrons 
in each monomer for treating the case accompanied with a charge moiety, 
we merely need to set the occupations to the corresponding number of electrons in each monomer.
\cite{gdf-eld}
Throughout this article, we treat cases with an overall singlet spin state 
in a spin-restricted manner.

Appendix \ref{practice-initial-dm} provides a practical example 
for preparing an initial density matrix within GD representation. 


\subsection{Time propagation of the density matrix in a GD representation}
\label{section-time-propagation}

In this study, the calculation of the time propagation of an electronic state is performed 
in terms of the Liouville--von Neumann equation associated with one particle density matrix 
as follows: 
\begin{align}
\dfrac{\partial}{\partial t}
\underline{\underline{\rho}}^\mathrm{GD} 
=
-\dfrac{i}{\hbar}
\left[
\underline{\underline{\mathcal{F}}}^\mathrm{GD}
\left\{ \underline{\underline{\rho}}^\mathrm{GD}(t),t \right\}, 
\underline{\underline{\rho}}^\mathrm{GD} 
\right], 
\label{VNLeq}
\end{align}
where 
\begin{align}
\underline{\underline{\mathcal{F}}}^\mathrm{GD}  
\equiv
\underline{\underline{F}}^\mathrm{GD}
 +
\underline{\underline{L}}^\mathrm{GD}.  
\end{align}
$\underline{\underline{L}}^\mathrm{GD}$ 
is a light--electron coupling matrix.
For the length gauge, their corresponding matrix elements have forms of 
$
\underline{\underline{L}}^\mathrm{GD}
=
+ e \underline{\underline{{\bf r}}}^\mathrm{GD} {\bf E} 
$, 
\cite{Faisal-LG-VG,ACP-laser}
where ${\bf E}$ is the three-dimensional electric field vector, 
which generally depend on a point in a three-dimensional space.  
We used the dipole approximation, namely, long wavelength approximation,  
\cite{Faisal-LG-VG,ACP-laser}
since the wavelength of light treated here is sufficiently large 
compared to the size of the molecular system treated. 

In the RT-TDDFT, the Fock matrix depends on the time-dependent density matrix.
For convenience of a discussion on the numerical demonstration 
to be presented in the later section, 
we label the dynamics described by Eq. (\ref{VNLeq}) including this dependency as 
'RT' meaning 'real time Fock matrix'
while the term 'FF' denoting 'frozen Fock matrix' is used to refer to 
the approximated dynamics with the replacement of the time dependent Fock 
by that at the initial simulation time.
For a technical simplicity and focusing on the projection scheme of electron dynamics, 
we employ the pure density functional for including an electronic exchange-correlation 
throughout this article.

For obtaining the time dependent electron density matrix, 
we solved the non-linear Liouville--von Neumann equation associated with 
the RT-TDDFT and introduced LP-GDF matrix 
by using the following two types of numerically stable time integrators, namely   
(1) the predictor-corrector second order Magnus scheme
with linear Fock extrapolation (PC2M-LF) 
\cite{Voorhis-P2CM-LF-PRB2006}
and 
(2) exponential propagation with predictor-corrector
SCF scheme using final corrector as a resultant density (EPPC1).
\cite{Herbert-EPPC-JCP2018}
PC2M-LF needs one time of update of Fock matrix for each step 
while an iteration scheme is applied for EPPC1 until a corrector density is converged.  
Their details are summarized in Appendix \ref{app-time-integrator}.  

We followed a dynamics with time steps of 8 and 20 attosecond 
using a PC2M-LF and EPPC1 time integrator, respectively.   
Hereinafter we abbreviate femtosecond to fs and attosecond to as.
In many cases with a moderate dynamics of density the use of former scheme is sufficient,   
while the latter was needed for a stable calculation of absorption spectrum 
using short laser pulses. 

The Fock, electron dipole transition matrices within the AO representation 
required for the dynamics calculation were evaluated 
by using the NTChem2013 software package.\cite{NTChem}

We should comment on conservations of properties during dynamics. 
As a special case, we examined a case starting from 
a locally excited state without external light field.
While within the LP-GD scheme a trace of density matrix is perfectly conserved 
both for RT-TDDFT and Frozen Fock approximation,
an expectation value of Fock operator is conserved only for the latter one.
Despite this, in a practical sense, this LP-GD RT-TDDFT scheme provides 
a good result with respect to a convergence to reference data 
as $\Delta_\epsilon$ increases, 
which will be shown in Sect. \ref{Numerical-demo}
with demonstrations of a userbility of the method
through calculations of charge migration dynamics and absorption spectrum. 
%

\section{Numerical demonstration}
\label{Numerical-demo}

Here, the size effect of the projected space on the electron dynamics 
and absorption spectrum are examined 
by using the LP-GDF electron dynamics method introduced in the previous section.  
We compare the time-dependent behaviors of the Mulliken charge of electron donor molecules 
for NPTL--TCNE with different initial excitation and continuum light field by varying $E_{\rm bw}$.
The convergence of absorption spectrum for 5UT is also examined in the same manner 
after showing explicitly the importance of the self-consistency between time-dependent density 
and the time-dependent Fock matrix during electron dynamics. 

\begin{figure}[th]
\includegraphics[width=0.5\textwidth]{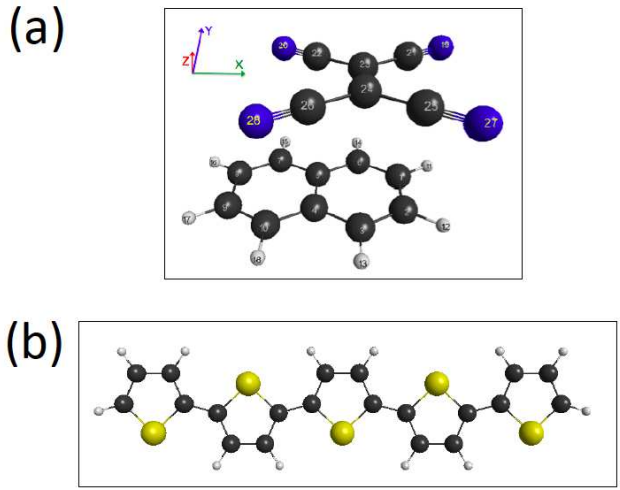}
\caption{Schematics of the geometries of the molecular aggregates treated in this article:  
(a) NPTL--TCNE dimer (b) 5UT
The centroids of aggregates were set to origin for all the systems. 
See the supplementary material for detailed information of their Cartesian coordinates. }
\label{Fig-1}
\end{figure}

%
With respect to the NPTL--TCNE, readers can find further information 
in our previous article.\cite{gdf-eld}
We validate the efficiency of the present projection method by showing 
a convergence of result with respect to a reduced size of local orbital space. 
The schematics of the molecular configurations 
used in this article as shown in Fig. \ref{Fig-1}. 

We also provide information related to the dependence of the computational cost 
in the electron dynamics calculations with and without Fock build 
on the size of the projection space, 
which are detailed in Appendix \ref{CPUtime-head-part}.

%
%

\subsection{NPTL--TCNE dimer}

As a first test system, we treat a dimer system consisting of NPTL and TCNE. 
In this combination of monomers, 
NPTL serves as an electron donor molecule,
while TCNE plays role as an electron acceptor. 
The molecular geometries and relative orientations of the monomers used here 
are the same as those in Panel (l) in Fig. 3 in our previous article \cite{gdf-eld}
on the original GDF electron dynamics scheme 
where readers can find further information including 
the literature of experimental data. 

The geometry of each monomer was optimized at the DFT/6-31G(d) level 
with the use of the PBE exchange correlation functional.\cite{PBE}  
The Fock matrix associated with initial optimized KS orbitals and its following 
time-dependent density matrix for the construction of the GDF matrix  
was also calculated at the same ab initio level. 
Both molecules have planar geometries in the optimized geometry 
in their ground electronic states. 
Here, as shown in Fig. \ref{Fig-1}, 
the principal axis of NPTL was set to be parallel to the X axis, 
while we set TCNE to be parallel to the Y axis. 
The molecular planes of these flat molecules are parallel to the X--Y plane. 
They were placed in a parallel orientation with a slide of 1.24 $\textrm{\AA}$ along the Y axis.
The distance between molecular planes was fixed at 4 $\textrm{\AA}$.  
Although a dimer is treated here, we considered 
the crystal data\cite{crystal-NTFL-TCNE-1967} reported in the literature 
with respect to the relative orientation. 
This selection of molecular configuration yields a non-vanishing overlap 
between the frontier orbitals, i.e., the HOMO of NPTL and the LUMO of TCNE.\cite{gdf-eld}
See the supplementary material for a further information 
on the geometrical coordinate of this system as well as GD and canonical orbital energies.

%
%


\begin{figure}[th]
\includegraphics[width=0.5\textwidth]{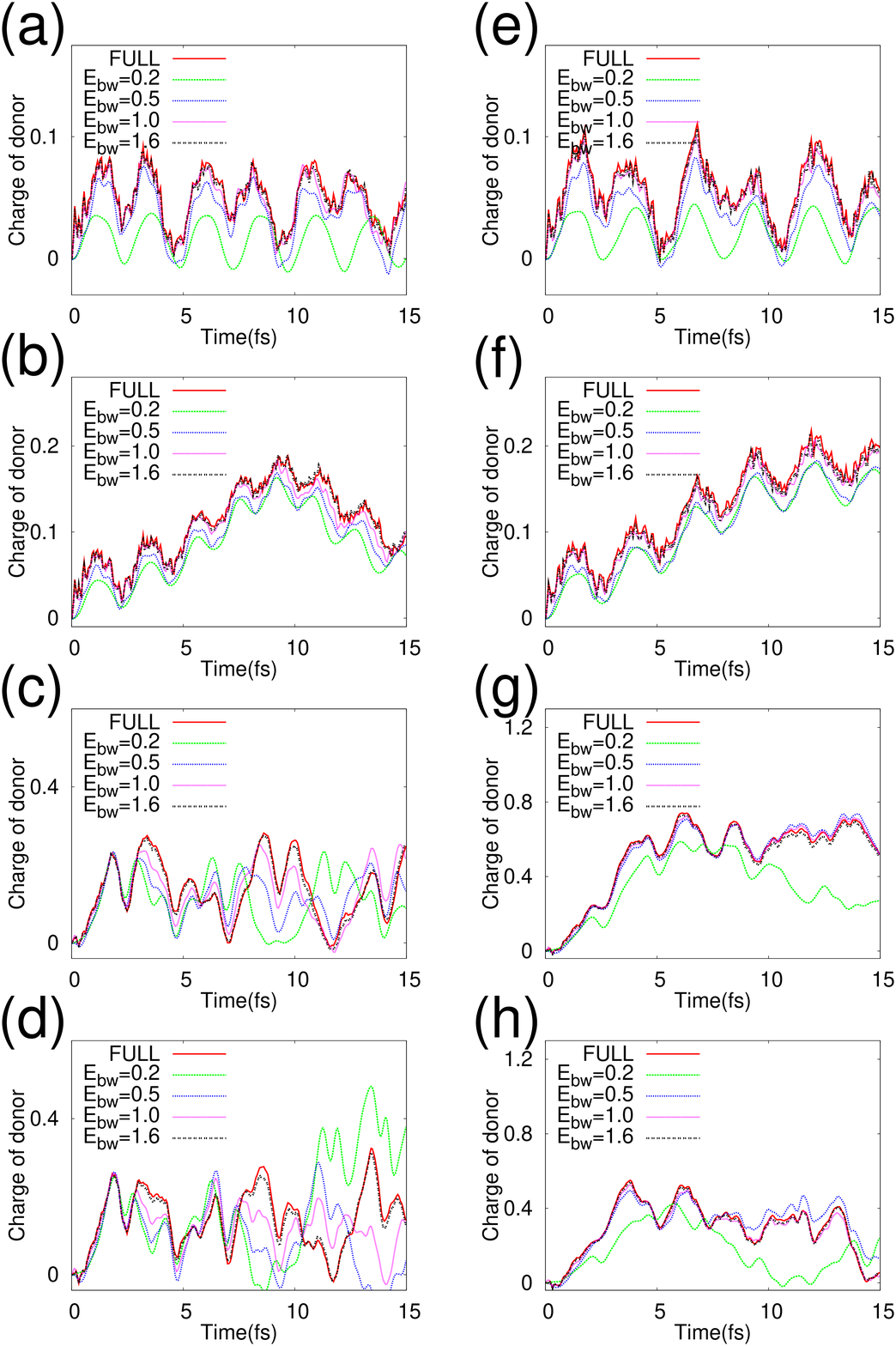}
\caption{
Convergence of the results for the charge separation dynamics in an NPTL--TCNE dimer
with an increase in the energy range determining the projected local orbital space. 
The energy width used for orbital projection is expressed by 
$E_{\rm bw}$ = $\Delta \epsilon$ in Hartree. 
The numbers of projected orbitals, $N_{\rm proj}$, are 13, 46, 74 and 120
respectively for E$_{bw}$ = 0.2, 0.5, 1.0 and 1.6.
The exact results obtained by using the full 296 orbitals are indicated by solid red line.
See the main text for the details of situations with respect to the treatment of Fock matrix,   
initial local excitation and light field employed in the panels. }
\label{Fig-2}
\end{figure}

Fig. \ref{Fig-2} provides the results of the excited electron dynamics 
involving initial local excitations and external light fields.
The time dependent behaviors of the Mulliken charge of the donor molecule, NPTL, 
are displayed with the variation of the energy range covering the projection orbital space 
$\Delta \epsilon$. In the panels of the figure, $\Delta \epsilon$ is expressed as 
$E_{\textrm{bw}}$. 
In panels, we plotted the Mulliken charge of NPTL as an electron donor system 
for each case corresponding to the vertical axis.
The horizontal axis denotes the time in femtoseconds.  

The results of RT-TDDFT LP-GDF scheme are displayed in panels of (a--d)  
while corresponding results of frozen Fock approximation are presented in (e--h). 
(a/e) and (b/f) show the results starting from the initial local excitation 
respectively in the donor NPTL and acceptor TCNE,  
while the other moieties are initially in the ground states within the GDF representation. 
In these four panels, no light field is irradiated to the system.  
In panels (c/g) and (d/h), a continuum light field is applied, and    
its field parameters as a wave length and unit vector of polarization are 700 nm 
corresponding to $\omega=0.065$ au 
and $\left( \frac{1}{\sqrt{3}}, \frac{1}{\sqrt{3}}, \frac{1}{\sqrt{3}} \right)$ 
in the XYZ Cartesian coordinate, respectively. 
The field strength $Es$ used in the cases of panels (c/g) is 0.015 a.u., 
while a larger strength of 0.02 is applied in (d/h).
The function form of light field is given in Appendix \ref{app-absorption}.
The continuum light field was replaced by a pulse 
with a sufficiently large time width 
$t_w=2.4\times10^{4}$ fs  
and peak time $t_c=0$. 


As seen in the RT cases (a) and (b) accompanied with initial local excitations 
without light fields show the fast convergence with respect to the increase of $E_{\textrm{bw}}$. 
In fact, $E_{\rm bw}=1.0$ associated with 74 projected local orbitals 
provides a quantitative convergence to the result obtained by the full 296 orbitals. 
On the contrary, in (c) and (d) from initial local ground states 
with continuum light field having moderate strength, the quantitative convergence are 
achieved by $E_{\textrm{bw}}$=1.6 providing 120 projected orbitals. 

As found in the panels (c/g) and (d/h), 
the FF approximation provides faster convergence compared to RT results, 
corresponding to the cases involved with light fields. 
On the other hand, the differences of the converged results of FF from those of RT 
means the importance of a consideration of feedback from time dependent density matrix 
to the Fock matrix.   
In (c/d/g/h), we employed the light stronger than that of usual sun light 
for a severe assessment of the present scheme in cases accompanied with hard excitations. 
In fact, in this strength, the non-linear effect about the density matrix 
result in the difference in convergence rates  
between frozen Fock type and RT-TDDFT calculations 
with respect to an increase of number of projected orbitals.

%
%

\begin{figure}[th]
\includegraphics[width=0.5\textwidth]{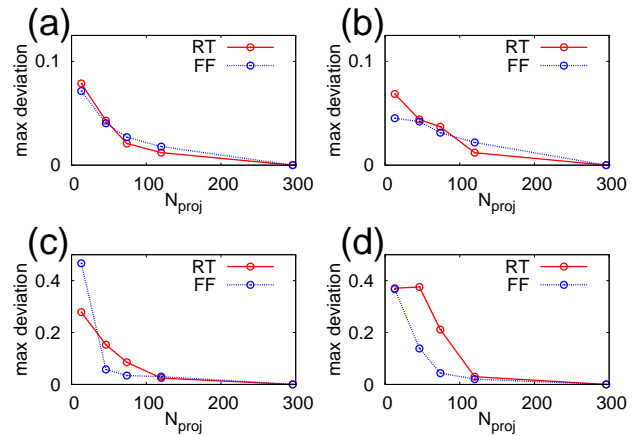}
\caption{$N_{\rm proj}$ dependency of max deviations for donnor molecule 
during dynamics from the results by full orbital calculation in Fig. \ref{Fig-2}.
Deviation is expressed as an absolute value but not relative one.  
The panels of (a), (b), (c) and (d) in this figure correspond to those of 
(a/e), (b/f), (c/g) and (d/h) in Fig. \ref{Fig-2}, respectively.  
``RT'' and ``FF'' in panels correspond to the results obtained by electron dynamics calculation 
using RT-TDDFT scheme considering the self-consistency and frozen Fock approximation.
}
\label{Fig-3}
\end{figure}

To clarify the convergence of time dependent behaviors of charge migration dynamics 
in Fig. \ref{Fig-2} with respect to the number of projected orbitals, 
we displayed the max deviations of the approximation results measured from 
the reference data obtained by full orbitals in Fig. \ref{Fig-3} 
as a function of $N_{\rm proj}$. 
As found in the figure, in all cases we can reduce a half the number of the orbitals 
keeping with the accuracy matched to the result obtained by using full orbitals.
The associated CPU times spent for the calculations and the ratios of them to 
those obtained by the full orbitals are summarized in Appendix \ref{CPUtime-head-part}.

%
 In the last part of this subsection, as a supplementary information, 
we provide an evidence of the efficiency of the orbital extraction scheme 
for the standard RT-TDDFT and Frozen Fock approximation 
by guiding to the corresponding results.  

 The electron dynamics calculation using the orbital extraction (projection) 
in standard KS orbital sequence is equivalent to 
the GDF electron dynamics calculation with $N_g=1$, namely no group division.
This is because a diagonalization of L\"{o}wdin AO representation in case of $N_g=1$ 
gives rise to exactly the same eigen states for the original Kohn-Sham equation 
at an initial simulation time. 
In the other words, in the case of $N_g=1$, total eigen states of Fock operator 
do not change through a transformation from a representation with original AO 
to that with L\"{o}wdin orthonormalized AOs. 

 We calculated charge migration of this system by treating the whole system as a monomer,
namely, by setting $N_g=1$.   
The energy threshold value for orbital extraction 
and number of selected orbitals were varied within LP scheme.  
The data were compared with those obtained by the methods without any orbital projection nor 
any group division, namely full orbital type of standard RT-TDDFT or Frozen Fock approach.
Because the setting of $N_g=1$ does not allow local excitation of each monomers at initial 
simulation time in the same sense as $N_g \ge 2$, 
we calculated only the light field cases starting from 
initial ground state of whole system, namely, no excitation of that. 
The results and detailed discussion are included in the Supplementary Information. 

We should also comment that the Frozen Fock approach without GD representation can provide 
a better convergence as seen in the comparison between the panel (g) in Fig. 2 and (c) in Fig. S2 
with respect to the $E_{\rm bw}=0.2$ and full orbital cases. 
In the same sense, RT-TDDFT without GD approach can be superior in convergence as found 
in the comparison between the panel (d) in Fig. 2 and (b) in Fig. S2 for 
$E_{\rm bw}=1.0$ and full orbital cases. 


\subsection{5UT}

In the second demonstration of the present work, 
we here apply the method to the calculation of absorption spectrum of 5UT. 
This molecule plays an important role not only as an electron donor but also 
a hole transfer material in an organic solar cell.\cite{polythiophene2017} 
The existence of sulfur atoms in $\pi$ conjugate system  
gives rise to the stability of molecular structure as well as 
the increase of charge conduction through the electron overlap 
contributed by their d-orbitals.\cite{Chung-polythiophene}
%
%
%
%
%
%
%
Here we do not discuss a chemical aspect of this system, 
and just focus on the performance of the method by looking 
the dependency of results on the size of projected local orbital space.

The molecular geometry of 5UT treated here was optimized 
in the PBE/6-31G(d) level calculation, 
which is commonly used in the electron dynamics calculation.  
The geometry information is summarized in the supplementary materials. 

The light pulse was applied and 
the time-dependent information of induced electronic dipole moments was converted 
to absorption spectrum, of which details on functional forms are given in Appendix 
\ref{app-absorption}. 
We used EPPC1 as the time integrator with the time step of 20as.    
The total time of simulation carried out was 30 fs. 
The functional form of light field is given in Appendix \ref{app-light}.
The field strength, $E_s$, central frequency, $\omega$, 
pulse peak time, $t_c$, pulse width time, $t_w$, and angular frequency, $\omega$,  
of the applied light pulse were 0.001au, 100(=2.42 fs), 40(=0.97 fs) and 0.0285, respectively.
The polarization vector of light field was set to be parallel to 
the vector of (1,1,1) in the Cartesian coordinate. We set the group number $N_g$=1.  
The initial density used in the electron dynamics was set to that of the DFT ground state.



\begin{figure}[th]
\includegraphics[width=0.5\textwidth]{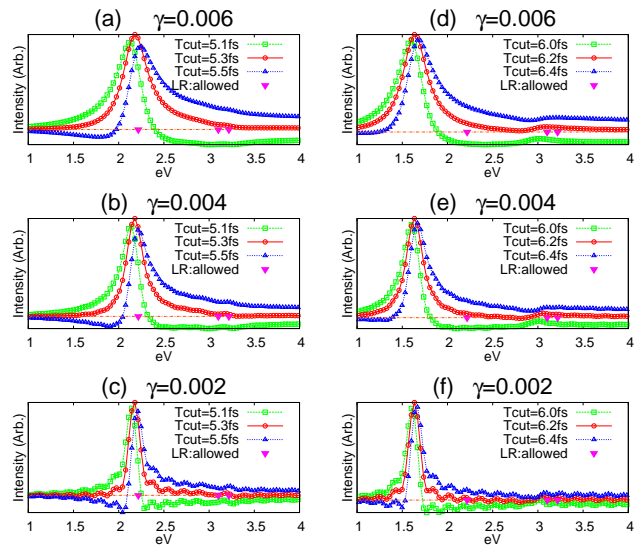}
\caption{The importance of feedback of time-dependent information to the Fock matix. 
We presented absorption spectrums of 5UT with the variation of the damping factor, 
$\gamma$, and time cut from the initial in the absorption calculation, $T_{\rm CUT}$. 
See also the supplementary materials for the mathematical form used for the absorption spectrum.
Left panels (a--c) and right ones (d--f) correspond to the results of RT and FF, respectively. 
Note that in all cases here, the full 394 orbitals are used in the dynamics calculations 
for preparing the light pulse induced electronic dipole moments for the spectrum calculations.
Triangles denotes the positions of optically allowed excitations 
associated with the absorption energy obtained by linear response TDDFT calculations.
}
\label{Fig-4}
\end{figure}

The absorption spectrum evaluated from the formula given in Appendix \ref{app-absorption}
numerically depends on the damping parameter, $\gamma$, and the initial cut off time, 
$T_{\rm CUT}$. 
We checked that the simulation time was long enough for the examined energy range from 0 to 4 eV.

%
%
%
%

In order to select a set of reliable parameters for absorption spectrum, 
we compared the obtained spectrum by varying $\gamma$ and $T_{\rm CUT}$, 
which is presented in Fig. \ref{Fig-4} including the information of positions 
of optically-allowed excitations in energy space obtained by LR-TDDFT.  
The transition properties evaluated by LR-TDDFT 
including excitation energies, oscillator strength, and dipole moments 
are summarized in the supplementary material. 
We also compared the results with and without the feedback of time-dependent density 
to the Fock matrix in order to show that the instantaneous back-reaction of 
the density to effective Hamiltonian in the simulation is the key factor 
for the reproduction of the LR-TDDFT results, 
which are correspondingly labeled with RT and FF defined 
in the section of theoretical method, Subsection \ref{section-time-propagation}.

With respect to the parameters, we found the best parameters $\gamma=0.004$ 
both for RT and FF cases in the aspects of the balance of 
proper smoothness and spectrum peak widths. In turn, 
the best parameters of $T_{\rm CUT}$ is 5.3 fs and 6.2 for RT and FF 
based on the observation of the appearance of maximum peaks.  
Again, we can see that RT calculations presented in the panels (a--c) correctly 
reproduce the first peak position given by LR-TDDFT while FF ones fail.
Then, we chose the RT type calculation for spectrum calculation 
with $\gamma=0.004$ and $T_{\rm CUT}$=5.3 fs 
and proceed to check the dependency on the size of projected local orbital space. 

%
%

\begin{figure}[th]
\includegraphics[width=0.4\textwidth]{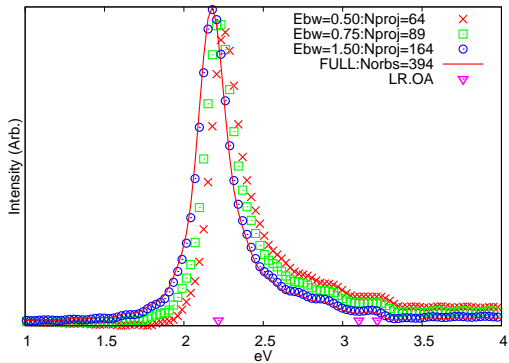}
\caption{Projected size dependency and convergence of absorption spectrum.  
we compared the results by projected local orbitals, $N_{\rm proj}$=64, 89, and 164 
correspondingly associated with $E_{\rm bw}$=0.5, 0.75, and 1.5 
to that obtained by using $N_{\rm orbs}$=394 full orbitals for 5UT molecule.  
Here we employed the damping factor of $\gamma=0.004$ and $T_{\rm CUT}$=5.3 fs
based on the observation in Fig. \ref{Fig-4}.
Triangles denotes the positions of optically allowed excitations associated with the 
absorption energy obtained by linear response TDDFT calculations, 
which are labeled with the symbol 'LR.OA'.}
\label{Fig-5}
\end{figure}

Fig. \ref{Fig-4} presents the convergence of the absorption spectrum 
using RT-TDDFT calculation combined with LP-GDF scheme 
with respect to the size of projected local orbital space.  
We can observe the monotonic convergence of the results by projection scheme
to that obtained by the 394 full orbitals. 
In fact, the spectrum shapes by $E_{\rm bw}$=0.5, 0.75 and 1.5 corresponding to 
$N_{\rm proj}$=64, 89 and 164 projected orbitals 
monotonically approaches to that of the full orbital calculation. 
More than half of the total orbitals were reduced in this spectrum calculation.
The information of CPU time is also included in the supplementary material. 
The success in the reduction of diabatic local orbital space 
required for the sufficient description 
suggests the efficiency of the introduced projection method for RT-TDDFT.


\section{Concluding remarks}
In this study, we introduced and assessed the electron dynamics method 
as a combination of the local orbital projection and GDF electron dynamics scheme
within a framework of RT-TDDFT.
Through the examination of complex charge migration induced by local excitation and light field  
for the NPTL--TCNE dimer as a test donor-acceptor system, we showed that 
the present method allows us to investigate the size of the Hilbert subspace 
associated with the excited electron dynamics of molecular aggregates. 
In the application of the projection method to absorption spectrum, 
we numerically demonstrated the reproduction of the result with 
use of the reduced number of orbitals. 

We can expect that this method paves a way to 
a practical investigation of electron dynamics of molecular aggregates 
with a further combination of the techniques for reducing the cost in Fock build.

We conclude by providing a perspective. 
The molecular motion, non-adiabatic mixing, and transition between 
electronic states can affect charge and exciton migration dynamics.    
%
%
There, a nuclear quantum effect may play an important role in electron dynamics 
in molecular aggregate systems having high density of states.
The GD representation scheme can afford a realistic and practical model Hamiltonian 
for treating these issues.
%
In our future work, we will report on a coarse grained quantum dynamics 
in molecular aggregates having a nano-size for describing a coupled dynamics 
of charge and exciton migration associated with their birth, 
extinction and transport.

\section{Supplementary material}
See supplementary material for the issues of  
(I) validity check of the time interval employed for describing the charge migration dynamics 
and 
(II) Efficiency of orbital projection method for standard RT-TDDFT and Frozen Fock method
     without a group division in charge migration dynamics for NPTL--TCNE system. 
(III) group diabatic and canonical orbital energies of NPTL--TCNE dimer
(IV) optical transition properties of 5UT obtained by LR-TDDFT calculation
(V) geometry data of molecular aggregates of a NPTL--TCNE dimer and 5UT.

\begin{acknowledgements} 
This research was supported by MEXT, Japan, 
``Next-Generation Supercomputer Project'' 
(the K computer project) 
and ``Priority Issue on Post-K Computer'' 
(Development of new fundamental technologies 
for high-efficiency energy creation, conversion/storage and use).
Some of the computations in the present study were performed 
using the Research Center for Computational Science, Okazaki, Japan,  
and also HOKUSAI system in RIKEN, Wako, Japan.  
%
%
\end{acknowledgements}

%
%

\noindent
\appendix

\section{Initial density matrix within GD representation associated with local excitations}
\label{practice-initial-dm}

Essential points in preparing initial density matrix are 
summarized in a following example.
Consider the spin-restricted case with three monomers 
each of which has 2 electrons and 3 orbitals. Here set $N_g=3$.
See the main text about the approximate treatment 
associated with this term of 'spin-restricted'. 
If we want to prepare the initial condition 
such that the first and third monomers are initially excited from local HOMO to LUMO,  
the initial density matrix in GDF scheme is constructed as follows: 
[1] set reference density matrix, namely GDF ground state, 
$\underline{\underline{\rho}}^{\textrm{GD:ground}}$
and   
[2] carry out local HOMO LUMO excitation and obtain the aimed GDF density, 
$\underline{\underline{\rho}}^{\textrm{GD:1,3-HL}}$ namely, 
\begin{align}
& \underline{\underline{\rho}}^{\textrm{GD:ground}} \equiv 
\left(
\begin{array}{ccccccccc}
2 & 0 & 0 & 0 & 0 & 0 & 0 & 0 & 0  \\ 
0 & 0 & 0 & 0 & 0 & 0 & 0 & 0 & 0  \\ 
0 & 0 & 0 & 0 & 0 & 0 & 0 & 0 & 0  \\ 
0 & 0 & 0 & 2 & 0 & 0 & 0 & 0 & 0  \\ 
0 & 0 & 0 & 0 & 0 & 0 & 0 & 0 & 0  \\ 
0 & 0 & 0 & 0 & 0 & 0 & 0 & 0 & 0  \\ 
0 & 0 & 0 & 0 & 0 & 0 & 2 & 0 & 0  \\ 
0 & 0 & 0 & 0 & 0 & 0 & 0 & 0 & 0  \\ 
0 & 0 & 0 & 0 & 0 & 0 & 0 & 0 & 0   
\end{array}
\right)  \\
\quad
& \Longrightarrow \notag \\
\quad
& \underline{\underline{\rho}}^{\textrm{GD:1,3-HL}} \equiv
\left(
\begin{array}{ccccccccc}
1 & 0 & 0 & 0 & 0 & 0 & 0 & 0 & 0  \\ 
0 & 1 & 0 & 0 & 0 & 0 & 0 & 0 & 0  \\ 
0 & 0 & 0 & 0 & 0 & 0 & 0 & 0 & 0  \\ 
0 & 0 & 0 & 2 & 0 & 0 & 0 & 0 & 0  \\ 
0 & 0 & 0 & 0 & 0 & 0 & 0 & 0 & 0  \\ 
0 & 0 & 0 & 0 & 0 & 0 & 0 & 0 & 0  \\ 
0 & 0 & 0 & 0 & 0 & 0 & 1 & 0 & 0  \\ 
0 & 0 & 0 & 0 & 0 & 0 & 0 & 1 & 0  \\ 
0 & 0 & 0 & 0 & 0 & 0 & 0 & 0 & 0   
\end{array}
\right).
\end{align}

\section{Time integrator}
\label{app-time-integrator}

We detail the time integrators used for solving non-linear Liouville von Neumann equation 
including time-dependent Hamiltonian depending on the density matrix employed in this study, 
namely, 
(1) the predictor-corrector second order Magnus scheme
with linear Fock extrapolation (PC2M-LF) 
and 
(2) exponential propagation with predictor-corrector
SCF scheme using final corrector as a resultant density (EPPC1).
Note again that PC2M-LF needs one time of update of Fock matrix for each step 
while EPPC1 includes an iteration scheme with respect to a convergence of corrector density. 
See the main text for the original articles of these schemes.

\subsection{PC2M-LF}
This scheme consists of the following five processes for each time step, 
\begin{align}
[1] &\, F_{3}  :=  - \frac{3}{4}F_{1a} + \frac{7}{4} F_{1b} \\
[2] &\, D_{4}  := 
    e^{- \frac{i}{\hbar} \frac{dt}{2} F_{3} } D_{2} e^{ \frac{i}{\hbar} \frac{dt}{2} F_{3} } 
\\
[3] &\, F_{5}  :=  F[D_{4}]  \\
[4] &\, D_{6}  := 
    e^{- \frac{i}{\hbar} dt F_{5} } D_{2} e^{ \frac{i}{\hbar} dt F_{5} } \\
[5] &\,( F_{1a}, F_{1b} ) :=  ( F_{1b}, F_{5} ) \,\, \Rightarrow \,\, \textrm{End of this step} 
\end{align}

If we consider a time propagation from $t$ to $t+dt$,  
$F_3$ and $F_5$ respectively denote the Fock matrices
at $t+\frac{1}{4}dt$ and $t+\frac{1}{2}dt$ 
while $D_2$, $D_4$ and $D_6$ correspond to the density matrices 
at $t$, $t+\frac{1}{2}dt$ and $t+dt$.
$F_3$ in the first step [1] is constructed from 
the Fock matrices, $F_{1a}$ and $F_{1b}$, respectively denote the Fock matrices
in the previous two times respectively at $t-\frac{1}{2}dt$ and $t-\frac{3}{2}dt$, 
by taking a linear extrapolation of them. 
It is known that owing to the time dividing points set properly   
this scheme has the same accuracy of second order Magnus expansion. 
The time propagation matrix having exponential form is treated exactly 
using a spectrum representation obtained by the diagonalization of Fock matrix, 
which is also applied in the EPPC1 scheme explained in the next subsection.



\subsection{EPPC1}

This integrator involved with a micro iteration 
consists of the following procedures: 
\begin{align}
[1] & \, F_{N}  :=   F[D_{N}] \\
[2] & \, D_{N+1}^{\rm pred}  := 
    e^{- \frac{i}{\hbar}  dt F_{N} } \, D_{N} \, e^{ \frac{i}{\hbar} dt F_{N} } \\
[3] & \, F^{\rm pred}_{N+1}  :=  F[D_{N+1}^{\rm pred}]  \\
[4] & \, F^{\rm mid}:=  \frac{1}{2} \left( F_{N} +  F_{N+1}^{\rm pred}  \right)  \\
[5] & \, D_{N+1}^{\rm corr}  := 
    e^{- \frac{i}{\hbar} dt F^{\rm mid} } \, D_{N} \, e^{ \frac{i}{\hbar} dt F^{\rm mid} } \\
[6] & \,\,  
\left\{
\begin{array}{l}
\textrm{If} \,\, || D_{N+1}^{\rm corr} - D_{N+1}^{\rm pred} || < \epsilon 
\,\,\, \\
\quad \textrm{then} \,\,\,
D_{N+1} := D_{N+1}^{\rm corr} 
\,\,\, \Rightarrow \,\,
\textrm{End of this step}  \\
\textrm{If} \,\, || D_{N+1}^{\rm corr} - D_{N+1}^{\rm pred} || \geq \epsilon 
\,\,\,  \\
\quad \textrm{then} \,\,\,
D_{N+1}^{\rm pred} := D_{N+1}^{\rm corr} 
\,\,\, \Rightarrow \,\,
\textrm{Return to [3]} 
\end{array} 
\right. 
\end{align}

Here, a time propagation is carried out 
from $t$ to $t+dt$, which are correspondingly labeled with $N$ and $N+1$. 
$F$ and $D$ are the Fock and density matrices.
'mid' denotes a mid point of $t$ and $t+dt$, namely, $t+\frac{1}{2}dt$. 
'pred' and 'corr' are abbreviations of 'predictor' and 'corrector', respectively.
The double '$||$' symbol means taking the Frobenius norm of a matrix 
placed between these two symbols. 
The threshold is expressed by $\epsilon=n\alpha\xi$ 
with $n$ being the dimension number of a corresponding matrix. 
$\alpha$ is set to be a value proportional to 
the absolute maximum eigen value of the matrix under consideration. 
Here, we used 1 as $\alpha$ for simplicity. 
Thus, $\xi$ determines the strictness of the self-consistency 
between the instantaneous density and the Fock matrix.

\section{Mathematical form of LP-GDF procedure}
\label{app-LP-GDF}

Here we provide the formal mathematical expression of LP-GDF scheme. 
At first, one knows that within the GD representation the identity operator is written as
\begin{align}
\hat{1}
\simeq 
\sum_{i=1}^{N_g} \sum_{j=1}^{N_{G_i}}
\mid
\phi_j^{G_i}
\rangle
\langle
\phi_{j}^{G_{i}}
\mid,
\end{align}
where 
$
\left\{
\mid \phi_j^{G_i} \rangle
\equiv 
\sum_k
\mid \widetilde{\chi}_{k,G_i} \rangle 
\left[ \underline{\underline{D}}_{G_i} \right]_{kj}
\right\}
$ 
with 
$ \mid \widetilde{\chi}_{k,G_i} \rangle $ being 
the k-th L\"{o}wdin orthogonalized atomic orbital basis function 
spanned in the G$_i$-th group 
is the set of GD localized orbitals 
for the group labeled by $G_i$ and $j$ ranges from 1 to $N_{G_i}$, 
which denotes the number of basis functions spanned at the site, $G_i$.
$i$ ranges from 1 to $N_g$, i.e., the number of monomer group sites.
We used the symbol for approximation in the equation 
because of the practical use of a finite basis set during computation.
Note that the GDF orbitals created from L\"{o}wdin orthonormal basis 
remain to be orthogonal under unitary transformations even if 
the transformations are carried out in each group. 
In turn, with respect to the structure of 
matrix representation of the Fock operator, 
only the GDF orbital pairs between different monomers are 
Fock non-orthogonal
while 
the pairs within the same group are 
Fock orthogonal
. 

Next, we introduce projection operators 
in order to realize steps (i)--(v) 
in Subsect.\ref{subsec-proj}
as follows:   
\begin{align}
\hat{P} \equiv
\sum_{i=1}^{N_g} \sum_{j \in \Omega_i}^{N_{G_i}}
\mid
\phi_j^{G_i}
\rangle
\langle
\phi_j^{G_i}
\mid
\end{align}
where 
\begin{align}
\Omega_i \equiv 
\left\{ 
j; 
| \epsilon_{j,G_i} - \bar{\epsilon} |
\le \Delta\epsilon_i/2 
\right\} 
\quad
\textrm{and}
\quad
\bar{\epsilon} \equiv (\epsilon_{\textrm{H}}+\epsilon_{\textrm{L}})/2. 
\end{align}
Here, $\epsilon_{j,G_i}$ is the local orbital energy 
associated with the GD orbital $\phi_j^{G_i}$, 
while $\epsilon_{\textrm{H}}$ and $\epsilon_{\textrm{L}}$
are the HOMO and LUMO energies of the whole system.

By using these mathematical tools within LP-GDF formulation, 
we approximate a one electron operator $\hat{O}$ as follows:  
\begin{align}
\hat{O}
\approx 
\hat{O}_P \equiv \hat{P}\hat{O}\hat{P}. 
\end{align}

\section{Function form of Laser pulse field}
\label{app-light}
The vector potential of light field as a function of time 
in the long wave length approximation employed takes a form of 
\begin{align}
{\bf A}(t) = \sum_j^{N_\textrm{p}} 
{\bf A}_{j} \, f(t;t_{c_j},t_{w_j}) \, \text{cos}(\omega_j (t-t_{c_j})+\delta_j),   
\end{align}
where bold font means the three dimensional vector in the Cartesian coordinate space. 
Here, the envelope function is defined by 
$ f(t;t_c,t_w) \equiv \textrm{exp} \left( - \left( \dfrac{t-t_c}{t_w} \right)^2 \right) $.  
The physical meanings of parameters appeared above are as follows; 
$t$ denotes time, $t_c$ is a field peak time, $t_w$ stands for a typical gaussian decay time, 
$\omega$ means a central angular frequency of field, and $\delta$ is a carrier envelope phase.
$N_\textrm{p}$ denotes a number of pulses. 
The electric field vector of external light corresponding to ${\bf A}(t)$
is given by $ {\bf E}(t) = -\dfrac{1}{c} \dfrac{\partial {\bf A}(t) }{\partial t} $. 
In the present article, we employed $N_\textrm{p}=1$ and $\delta=0$. 


\section{Absorption spectrum}
\label{app-absorption}

The absorption spectrum was evaluated using the following function,   
\begin{align}
S(\omega) 
\equiv
\dfrac{1}{3}\textrm{Tr}[\underline{\underline{\widetilde{\sigma}}}(\omega)]
,
\end{align}
where
$
\underline{\widetilde{\underline{\sigma}}}(\omega)
\equiv
\dfrac{4\pi\omega}{c}\textrm{Im}[\underline{\underline{\widetilde{\alpha}}}(\omega)]
$
is an absorption cross section with
$
[\underline{\underline{\widetilde{\alpha}}}(\omega)]_{jk}
\equiv
\dfrac{\widetilde{\mu}^{\rm{ind}}_j(\omega)}{\widetilde{E}_k(\omega)}
=
\dfrac
{ \int_{T_{\rm CUT}}^{T_{\rm FIN}} e^{ i \omega t } e^{ - \gamma t} \mu_{j}(t) }
{ \int_{T_{\rm CUT}}^{T_{\rm FIN}} e^{ i \omega t } E_k(t) }
$. 
$\gamma$ is a damping factor. 
Here, 
$ \mu^{\rm{ind}}_j(t) \equiv \mu_j(t) - \mu_j(0) $ and ${E_k(t)}$ 
are induced dipole moment and external electric field, respectively. 
The symbol of tilde means taking the Fourier transformation.
$T_{\rm FIN}$ is a final simulation time. 
$T_{\rm CUT}$ is the initial time used for a transformation 
from time dependent induced dipole moment of electrons and external field 
to absorption spectrum.

\section{Computational time}
\label{CPUtime-head-part}

Here, we provide information about the computational cost of electron dynamics 
including analysis of the time-dependent properties. 
Note that the overhead of parallel computation is also included in time. 
The aim here is to show examples of the performance of the method   
in cases with use of moderate computer facilities.  


The specifications of the computer and compiler utilized in this article are as follows:
[For NPTL--TCNE]
\{ Intel(R) Xeon(R) Gold 6148 0 @ 2.40GHz with a cache size of 27.5MB,   
  The Intel FORTRAN compiler in Version 18.0.2.199 with level three optimization.
\}
[For 5UT]
\{
  Intel(R) Xeon(R) Gold 6152 0 @ 2.10GHz with a cache size of 30.9MB,   
  The Intel FORTRAN compiler in Version 17.0.4.196 with level three optimization.
\}

In all the calculations 40 cpu cores were used in a parallel manner 
using openMP for matrix-matrix product and MPI for Fock build if needed.

\subsection{NPTL--TCNE}
\label{CPUtime-NTPL-TCNE}

Tab. \ref{dyn-analy-cost-NapTcne-PC2M} and Tab. \ref{dyn-analy-cost-NapTcne-EPPC1}
present the comparison of the computation times using different number of projected local orbitals 
in cases of PC2M-LF and EPPC1 methods, respectively. The system treated is the NPTL--TCNE dimer.
A ratio to the time spent in cases with full orbitals expresses an acceleration in computation 
by using projected space. 
The tables include the comparisons between RT and FF. 
Note that the time intervals of one time step employed here are different 
between these two integrators. 
We checked the reproduction of the same result by using these two integrators, 
which is presented in the supplementary material.

In Tab. \ref{dyn-analy-cost-NapTcne-PC2M} for PC2M-LF,  
though we can see that a moderate acceleration is attained with a help of projected space 
and acceleration in FF is superior to RT one including Fock build, 
the use of RT recommended for the reliable calculation 
and provides still moderate acceleration by projections of local orbital space.  
In fact, the computation time in RT with $E_{\rm bw}=1.6$ corresponding to $N_{\rm proj}=120$ 
which provide converged results is reduced to one fourth of that 
spent in the cases using the full orbitals.  

As seen in Tab. \ref{dyn-analy-cost-NapTcne-EPPC1},   
the acceleration in the calculation by projection in case of EPPC1  
including micro iterations accompanied with Fock build for a convergence of corrector density 
is less than those of PC2M-LF in Tab. \ref{dyn-analy-cost-NapTcne-PC2M}. 
Despite of this, the case of $E_{\rm bw}=1.6$ with $N_{\rm proj}=120$ 
need approximately the half of the computational time in the full orbital case.

\begin{table}[t]
\begin{center}%
\begin{tabular}{cccc}
\hline
   $E_{\rm bw}$  &\quad  $N_{\rm proj}$ &\quad  RT &\quad  FF  \\
\hline
  0.2  &\quad  13  &\quad 108[0.14]$\{1\}$  &\quad  16[0.03]$\{0\}$   \\
  0.5  &\quad  46  &\quad 120[0.15]$\{1\}$  &\quad  24[0.04]$\{0\}$   \\
  1.0  &\quad  74  &\quad 141[0.18]$\{1\}$  &\quad  40[0.06]$\{0\}$   \\
  1.6  &\quad 120  &\quad 203[0.26]$\{1\}$  &\quad  93[0.14]$\{0\}$   \\
  FULL &\quad 296  &\quad 801[1.00]$\{1\}$  &\quad 679[1.00]$\{0\}$   \\
\hline
\end{tabular}
\end{center}
\caption{CPU time spent for the dynamics calculations plus the time-dependent analysis 
for NPTL--TCNE with PC2M-LF scheme with 1875 time steps using 8 as time interval.
The unit of time is second.
Numeric value in [] denotes the ratio to the cost in case with use of full orbitals
while that in $\{\}$ mean the number of Fock build per one time step.
Note the single Fock build per one time step in the PC2M-LF scheme. 
The spent time needed in the cases of Fig. \ref{Fig-2} are the same 
in each group of the panels (a-d) for RT and (e-h) for FF, namely, 
in the above RT and FF columns correspond to the group of (a-d) and (e-h) in Fig. \ref{Fig-2}.
See the text for the meanings of $E_{\rm bw}=\Delta \epsilon$, $N_{\rm proj}$, RT and FF.
The wall time means the wall clock time associated with 
openMP for dynamics and MPI for Fock build.}
\label{dyn-analy-cost-NapTcne-PC2M}
\end{table}

\onecolumngrid

\begin{table}[b]
\begin{center}%
\begin{tabular}{cccccc}
\hline
   $E_{\rm bw}$  &  $N_{\rm proj}$ & (a) & (b) & (c) & (d) \\
\hline
  0.2  &  13  & 185[0.17]$\{2.6\}$  &  155[0.14]$\{2.0\}$   &  229[0.20]$\{3.1\}$   &  256[0.21]$\{3.4\}$   \\
  0.5  &  46  & 239[0.21]$\{3.5\}$  &  267[0.24]$\{4.0\}$   &  310[0.26]$\{4.0\}$   &  332[0.27]$\{4.0\}$   \\
  1.0  &  74  & 302[0.27]$\{4.0\}$  &  312[0.28]$\{4.2\}$   &  350[0.29]$\{4.1\}$   &  360[0.29]$\{4.1\}$   \\
  1.6  & 120  & 621[0.56]$\{5.0\}$  &  457[0.41]$\{5.0\}$   &  534[0.45]$\{5.0\}$   &  545[0.44]$\{5.0\}$   \\
  FULL & 296  & 1121[1.00]$\{6.1\}$  & 1132[1.00]$\{6.2\}$  &  1205[1.00]$\{6.2\}$  &  1256[1.00]$\{6.7\}$   \\
\hline
\hline
   $E_{\rm bw}$  &  $N_{\rm proj}$ & (e) & (f) & (g)  & (h) \\
\hline
  0.2  &  13  &  8[0.04]$\{1\}$  &  8[0.04]$\{1\}$   &  8[0.03]$\{1\}$   &  8[0.03]$\{1\}$    \\
  0.5  &  46  & 11[0.05]$\{1\}$  & 11[0.05]$\{1\}$   & 11[0.04]$\{1\}$   & 11[0.04]$\{1\}$    \\
  1.0  &  74  & 17[0.07]$\{1\}$  & 17[0.07]$\{1\}$   & 17[0.07]$\{1\}$   & 17[0.07]$\{1\}$    \\
  1.6  & 120  & 38[0.14]$\{1\}$  & 39[0.15]$\{1\}$   & 39[0.14]$\{1\}$   & 38[0.14]$\{1\}$    \\
  FULL & 296  & 274[1.00]$\{1\}$ & 271[1.00]$\{1\}$  & 283[1.00]$\{1\}$  & 283[1.00]$\{1\}$    \\
\hline
\end{tabular}
\end{center}
\caption{CPU time spent for the dynamics calculations plus the time-dependent analysis 
for NPTL--TCNE with EPPC1 scheme with 750 time steps using 20 as time interval.
The unit of time is second.
Numeric value in [] denotes the ratio to the cost in case with use of full orbitals 
while that presented in $\{\}$ correspond to the average number of cycles in  
micro-iterations needed for the density convergence with $\xi=10^{-8}$. 
See the text for the meanings of $E_{\rm bw}=\Delta \epsilon$, $N_{\rm proj}$, RT and FF.
(a-d) and (e-h) correspond to RT and FF, respectively, 
and all the symbols are the same as Fig. \ref{Fig-2} in the main text. 
The definition of the wall time is the same as that in Tab. \ref{dyn-analy-cost-NapTcne-PC2M}.
}
\label{dyn-analy-cost-NapTcne-EPPC1}
\end{table}

\twocolumngrid


\subsection{5UT}
\label{CPUtime-5UT}

Tab. \ref{dyn-analy-cost-5UT-EPPC1} presents the comparison of the computation times 
using different number of projected local orbitals in cases of EPPC1 methods for 5UT system.
The way of presentation is the same as the table for NPTL--TCNE. 
Here also we show the comparisons between RT and FF. 
As seen in Tab. \ref{dyn-analy-cost-5UT-EPPC1}, 
the case of $E_{\rm bw}=1.5$ with $N_{\rm proj}=164$ associated 
with the converged spectrum presented in Fig. 5 in the main text 
needs merely less than half of the computational time in the case with the full 394 orbitals.

\begin{table}[t]
\begin{center}%
\begin{tabular}{cccccc}
\hline
   $E_{\rm bw}$  &\quad  $N_{\rm proj}$ &\quad CPUtimes \\
\hline
  0.5  &\quad  64  &\quad  51949[0.43]$\{2.8\}$  \\
  0.75 &\quad  89  &\quad  53447[0.44]$\{2.9\}$  \\
  1.5  &\quad 164  &\quad  61075[0.50]$\{2.9\}$  \\
  FULL &\quad 394  &\quad 123033[1.00]$\{3.8\}$ \\
\hline
\end{tabular}
\end{center}
\caption{CPU time spent for the dynamics calculations plus the time-dependent analysis 
for 5UT with EPPC1 scheme with 1500 time steps using 20as time steps.
The unit of time is second.
Numeric value in [] denotes the ratio to the cost in case with use of full orbitals  
while that presented in $\{\}$ correspond to the average number of cycles in  
micro-iterations needed for the density convergence with $\xi=10^{-8}$. 
See the text for the meanings of $E_{\rm bw}=\Delta \epsilon$ and $N_{\rm proj}$.
The wall time means the wall clock time, and the values in parentheses 
denote the CPU efficiency in percent associated openMP for dynamics and MPI for Fock build.}
\label{dyn-analy-cost-5UT-EPPC1}
\end{table}

\clearpage


\clearpage

\renewcommand{\tablename}{TABLE.S}
\renewcommand{\figurename}{FIG.S}

\renewcommand{\appendixname}{}

{\bf
\large 
Supplementary material for ``Electron dynamics method using a locally projected group diabatic Fock matrix for molecule and aggregate'' 
\normalsize
}

\renewcommand{\thesection}{\Roman{section}}
\def\thesection{\Roman{section}}

\setcounter{section}{0}
\setcounter{figure}{0}
\setcounter{table}{0}

\section{Convergence of charge migration dynamics with respect 
to the time increment using different integrator for NPTL--TCNE system.}

\begin{figure}[b]
\includegraphics[width=0.55\textwidth]{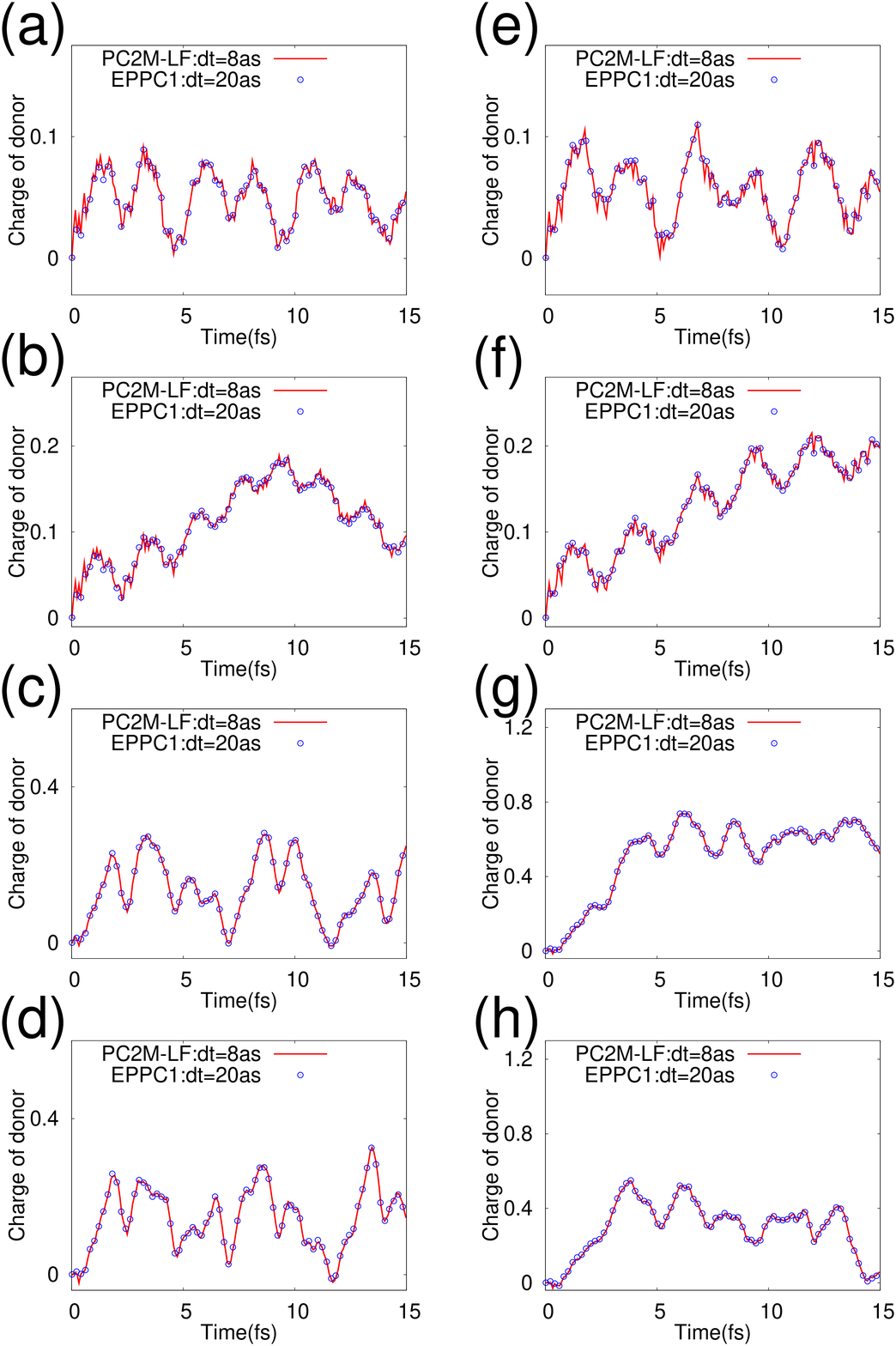}
\caption{Check of the convergence of charge migration dynamics with respect 
to the time increment using different integrator for NPTL--TCNE system.
See the main text in this supplementary material about the panels.}
\label{SFig-1}
\end{figure}

Fig. S\ref{SFig-1} shows the convergence of charge migration dynamics 
for NPTL--TCNE system with respect to time increments.  
The panels correspond to those in Fig. 2 in the main article.
For a stringent check, we used different integrators, namely, PC2M-LF and EPPC1, 
with the time increments of 8 and 20 as in each step, respectively. 
The perfect agreement with each other means the convergence of the result.

\clearpage

%
%

\section{
Efficiency of orbital projection method for standard RT-TDDFT and Frozen Fock method 
without a group division in charge migration dynamics for NPTL--TCNE system.}

We demonstrate an efficiency of the present local projection (LP) scheme 
for a standard RT-TDDFT. Note that the standard RT-TDDFT is equivalent 
to the case with $N_g$=1 and a ground state of whole system as an initial state 
within LP-GDF electron dynamics scheme.  

This is also verified for the calculation of absorption spectrum for 5UT with $N_g$=1,   
which is presented in Fig. 5 in the main text.  
Here, for charge migration dynamics in NPTL--TCNE system,  
we show the efficiency of LP scheme for the standard RT-TDDFT 
without any group division in the Fock matrix.
We also show the results for the Frozen Fock approach. 

\begin{figure}[b]
\includegraphics[width=0.53\textwidth]{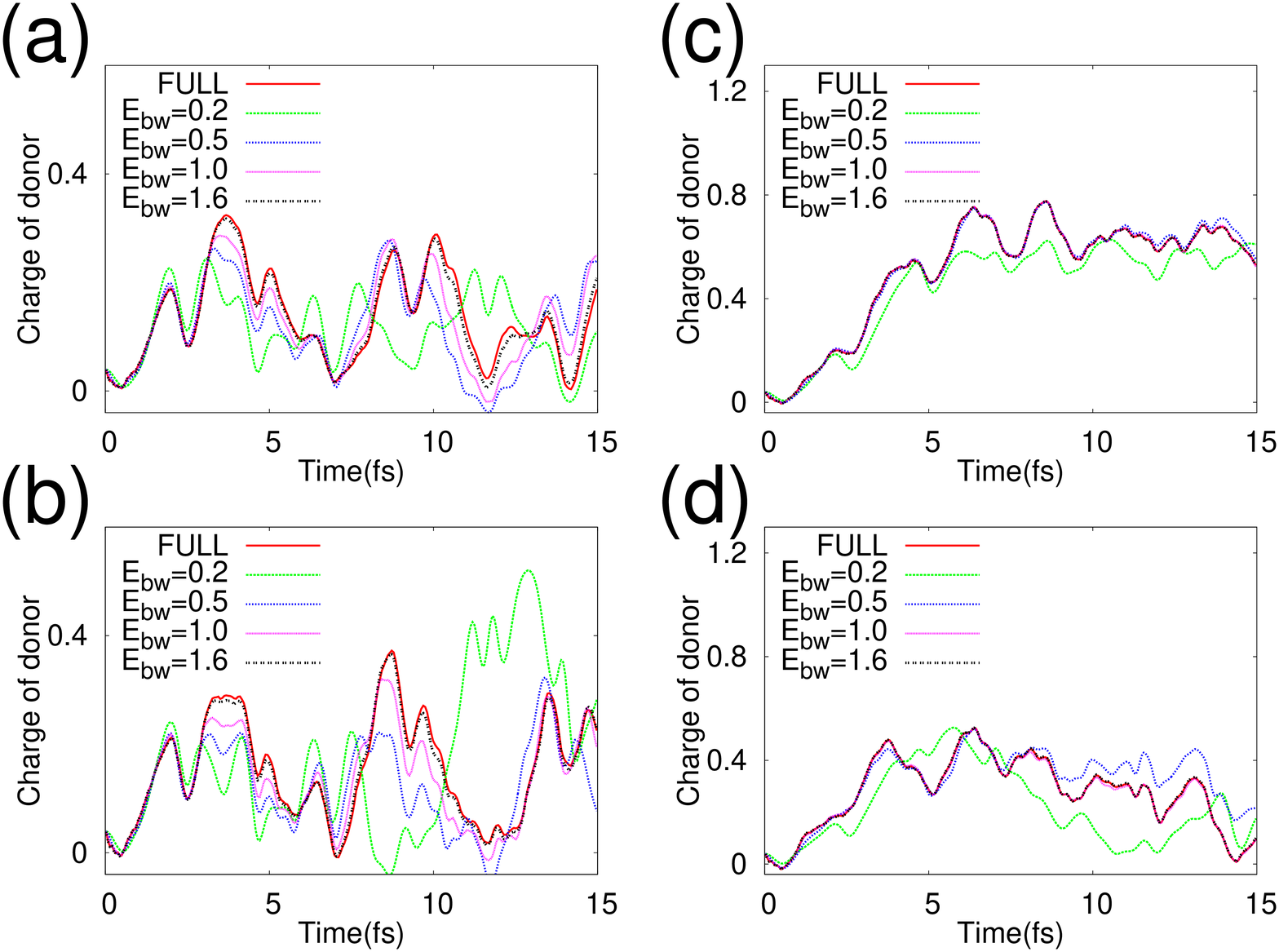}
\caption{
Light induced charge migration dynamics with $N_g$=1 for NPTL--TCNE.
Panels (a), (b), (c) and (d) in this figure 
respectively correspond to (c), (d), (g) and (h) in Fig. 2 in the main text.
}
\label{SFig-2}
\end{figure}

In this context, 
panels (a/b) and (c/d) in Fig. S\ref{SFig-2} with $N_g$=1 in LP-GDF electron dynamics 
correspond to (c/d) and (g/h) of Fig. 2 in the main text with $N_g$=2, respectively. 
The same properties are plotted. Time integrator employed here is EPPC1. 
Strictly speaking, the initial electronic condition is different for both schemes 
since the former standard case starts from the initial density matrix made from 
the ground state of whole system while the initial density matrix was prepared by using 
the combination of local ground states of monomers in the latter LP-GDF case. 
Despite this, the time dependent behavior is similar for both cases, 
which inversely supports validity of group division in Fig. 2.

\begin{figure}[b]
\includegraphics[width=0.35\textwidth]{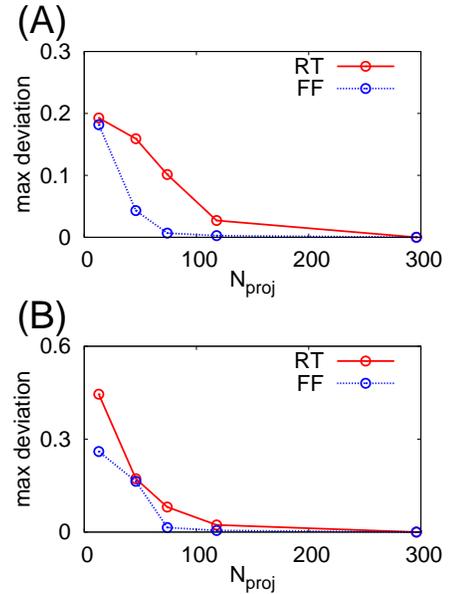}
\caption{
$N_{\textrm{proj}}$ dependency of max deviations for donor
molecule during dynamics from the results by full orbital calculation
in Fig. S\ref{SFig-2}.
RT and FF in panel (A) were evaluated from the results 
in (a) and (c) of Fig. S\ref{SFig-2}. 
while RT and FF in panel (B) were evaluated from the results in (b) and (d).
}
\label{SFig-3}
\end{figure}

Fig. S\ref{SFig-3} is the maximum error analysis for the LP results 
compared from the full orbital calculations. 
Panels (A) and (B) correspond to the cases of (a/b) and (c/d) in Fig. S\ref{SFig-2} 
The plots of the data were carried out in the same way as Fig. 3 in the main text.

These results indicate that 
the LP scheme can safely reduce the computational cost with respect to orbital numbers 
required for the description also for charge migration dynamics under a light field 
with using standard RT-TDDFT equivalent to LP-GDF-electron dynamics scheme with $N_g$=1.

\color{black}

\clearpage

\section{Group diabatic and canonical orbital energies of NPTL--TCNE dimer}

Tab. \ref{GDLO-energy-nap-tcne} 
summarizes the energy differences of LUMO+1, LUMO and HOMO-1 measured from HOMO,   
which are presented for group diabatic and canonical representations 
associated with monomers and dimer system, respectively.

\begin{table}[b]
\begin{center}%

\begin{tabular}{|l|c|c|c|}
\hline 
GD &  (D)NPTL & (A)TCNE \\
\hline 
$\Delta\epsilon^{\rm G}_{{\rm L}+1,{\rm H}} $  &$+0.1554$  & $+0.1925$   \\
$\Delta\epsilon^{\rm G}_{{\rm L},{\rm H}}   $  &$+0.1244$  & $+0.0939$   \\  
$\Delta\epsilon^{\rm G}_{{\rm H},1-{\rm H}} $  &$-0.0312$  & $-0.0182$   \\
\hline
\end{tabular}
\quad \\
\quad \\
\begin{tabular}{|l|c|c|}
\hline 
CA    &  NPTL--TCNE \\
\hline 
$\Delta\epsilon^{\rm W}_{{\rm L}+1,{\rm H}} $  &  $+0.1247 $ \\
$\Delta\epsilon^{\rm W}_{{\rm L},{\rm H}} $     &  $+0.0586 $ \\
$\Delta\epsilon^{\rm W}_{{\rm H}-1,{\rm H}} $   &  $-0.0303 $ \\
\hline 
\end{tabular}
\end{center}
\caption{
(Upper table) 
Energy differences of group diabatized HOMO$-1$, LUMO, and LUMO$+1$ measured 
from group diabatized HOMO of the electron donor NPTL and acceptor TCNE moieties 
in a molecular complex, which are labeled with the symbols of 
$\Delta\epsilon^{\rm G}_{{\rm H}-1,{\rm H}}$, 
$\Delta\epsilon^{\rm G}_{{\rm L},{\rm H}} $, and 
$\Delta\epsilon^{\rm G}_{{\rm L}+1,{\rm H}} $, respectively. 
Note that the monomer interactions are included 
in the evaluations of the local orbital energies of monomers.
(Lower table)
Energy differences of canonical HOMO$-1$, LUMO and LUMO$+1$ measured from HOMO 
for the whole NPTL--TCNE dimer system, which are denoted by the symbols of 
$\Delta\epsilon^{\rm W}_{{\rm H}-1,{\rm H}}$, 
$\Delta\epsilon^{\rm W}_{{\rm L},{\rm H}} $, and 
$\Delta\epsilon^{\rm W}_{{\rm L}+1,{\rm H}} $, respectively. 
In both tables, the unit is atomic unit.
GD and CA mean 'group diabatic' and 'canonical', respectively.
}
\label{GDLO-energy-nap-tcne}
\end{table}


\section{Excitation energies, transition dipoles and oscillator strength of 5UT}

Tab.S\ref{5UT-LR-TDDFT} summarizes the information on lower excitations of 5UT 
at the geometry given in the later section in this supplementary material. 
We showed the excitation energies, transition dipole moment vectors, and oscillator strength. 

Here, $(i \rightarrow j)$, $\Delta E_{ij}\equiv|E_i-E_j|$, $\vec{L}_{ij}$ and 
$f_{ij} \equiv \frac{2}{3}(\Delta E_{ij}) |\vec{L}_{ij}|^2 $
denote a state pairs of transition, energy difference of them, 
transition dipole moment vector for them, and its corresponding oscillator strength, respectively.
$E_i$ means $i$-th adiabatic state with 0 being ground electronic state.
The geometry of this molecular system is given later in this supplementary material. 
According to magnitudes of oscillator strength, we considered that 
0$\rightarrow$1, 0$\rightarrow$3 and 0$\rightarrow$5 are optically allowed transitions, 
and used as reference excitation energies in the presentation of Fig. 4 and 5 in the main text. 

Though it is not the aim to reproduce the experimental data in the present article 
since the main focus is the projection scheme in electron dynamics calculation, 
we provide information near to the situation treated here with respect to the target system. 
%
%
The experimental data of the position of the first peak in absorption spectrum of  
optical electronic transition for this 5UT system, however, in the CHCl$_3$ solvent 
but not in a gas phase, is reported in the former works, 
\cite{spec-5UT-in-CHCl3}
and its value is 2.98 eV.  
Though the absorption spectrum for electronic transition in a molecular system 
can be changed by an existence of a polar solvent associated with 
a complex modification of electric transition and structure relaxation,
this is not significant in this system. 

On the other hand, as seen in Fig. 4 of the main article as the comparison part of 
RT and FF results, 
the value given by RT-TDDFT, 2.2 eV, being consistent with LR-TDDFT result here  
within the present limited ab initio level is 
close to the experimental value mentioned above, 
compared to the peak value, 1.6 eV, given by FF approximation as found in Fig. 4. 
This observation also supports the basic assertion of 
the importance of the self-consistent treatment for the Fock matrix and 
time-dependent density matrix in a calculation of absorption spectrum. 

The remaining discrepancy in the LR-TDDFT result here is remedied by using 
more appropriate higher level basis set\cite{spec-5UT-in-CHCl3}
and exchange-correlation functionals \cite{polythiophene2017}.

In fact, at the same structure, 
the use of the representative hybrid functionals with a long-range correction, namely, 
CAM-B3LYP\cite{CAM-B3LYP}
and LC-BOP\cite{LC-BOP},
provide the optically allowed first electronic transition energies in gas phase, 
2.80 eV and 2.99, respectively, while their corresponding values for 
the polarizable continuum model(PCM)\cite{PCM}
with CHCl$_3$ solvent at the same structure are 2.68 and 2.88. 
In the calculations using the PCM combined with LR-TDDFT here, 
we used the ab initio program package 
for electronic structure calculation, GAMESS.\cite{GAMESS} 

%


%
%
%
%

\begin{table}[b]
\begin{tabular}{|c|c|c|c|c|c|}
\hline 
$( i \rightarrow j )$ & $\Delta E_{ij}$(eV) & \multicolumn{3}{|c|}{$\vec{L}_{ij}$(au)}  & $f_{ij}$(au) \\
\hline 
\hline 
0 $\rightarrow$ 1   &   2.22  &   $+$5.11   &   $+$0.00   &  $+$0.00  &  1.42 \\
0 $\rightarrow$ 2   &   2.37  &   $+$0.00   &   $-$0.01   &  $+$0.00  &  0.00 \\
0 $\rightarrow$ 3   &   3.10  &   $-$1.16   &   $+$0.00   &  $+$0.00  &  0.10 \\
0 $\rightarrow$ 4   &   3.18  &   $+$0.00   &   $-$0.13   &  $+$0.00  &  0.00 \\
0 $\rightarrow$ 5   &   3.22  &   $+$1.64   &   $-$0.00   &  $+$0.00  &  0.21 \\
0 $\rightarrow$ 6   &   3.57  &   $+$0.00   &   $-$0.06   &  $+$0.00  &  0.00 \\
0 $\rightarrow$ 7   &   3.59  &   $+$0.00   &   $+$0.00   &  $+$0.00  &  0.00 \\
0 $\rightarrow$ 8   &   3.70  &   $+$0.00   &   $-$0.09   &  $+$0.00  &  0.00 \\
0 $\rightarrow$ 9   &   3.80  &   $-$0.25   &   $-$0.00   &  $+$0.00  &  0.01 \\
0 $\rightarrow$ 10  &   3.83  &   $+$0.00   &   $+$0.14   &  $+$0.00  &  0.00 \\
\hline 
\end{tabular}
\caption{
Transition properties of 5UT obtained by LR-TDDFT calculation with PBE/6-31G(d).
See the main text in this supplementary material for the details.}
\label{5UT-LR-TDDFT}
\end{table}

\clearpage

\section{Geometry data of molecular systems}
We summarize the data of Cartesian coordinates in Angstrom for the molecular systems 
used in the present article. The information of atoms are also included below.

\subsection{NPTL--TCNE}
\begin{verbatim}
C     1.824214    0.710474   -1.999379
C     1.824214   -0.710473   -1.999379
C     0.628132   -1.408746   -1.999379
C    -0.619806   -0.721697   -1.999379
C    -0.619809    0.721696   -1.999379
C     0.628130    1.408742   -1.999379
C    -1.867746    1.408746   -1.999379
C    -3.063829    0.710474   -1.999379
C    -3.063829   -0.710473   -1.999379
C    -1.867746   -1.408742   -1.999379
H     2.775875    1.251891   -1.999379
H     2.775876   -1.251890   -1.999379
H     0.624772   -2.504937   -1.999379
H     0.624771    2.504934   -1.999379
H    -1.864388    2.504937   -1.999379
H    -4.015491    1.251889   -1.999379
H    -4.015490   -1.251891   -1.999379
H    -1.864388   -2.504934   -1.999379
N     2.832029    2.070410    2.000621
N    -1.591624    2.070452    2.000621
C     1.839916    1.437817    2.000621
C    -0.599565    1.437776    2.000621
C     0.620188    0.694253    2.000621
C     0.620194   -0.694285    2.000621
C     1.839957   -1.437800    2.000621
C    -0.599550   -1.437827    2.000621
N     2.832015   -2.070408    2.000621
N    -1.591638   -2.070398    2.000621
\end{verbatim}

\subsection{5UT}

\begin{verbatim}
C    -8.572000    1.432812    0.000000
C    -9.102000    0.159612    0.000000
S    -7.858600   -1.051488    0.000000
C    -6.587100    0.165912    0.000000
C    -7.149500    1.439512    0.000000
H    -9.185200    2.337012    0.000000
H   -10.148900   -0.142688    0.000000
H    -6.546000    2.351012    0.000000
C    -4.632800   -1.495088    0.000000
C    -5.196000   -0.220788    0.000000
S    -3.927900    0.998512    0.000000
C    -2.656500   -0.219888    0.000000
C    -3.219600   -1.495488    0.000000
H    -5.236800   -2.406288    0.000000
H    -2.615500   -2.406588    0.000000
C    -0.706100    1.441412    0.000000
C    -1.270100    0.165512    0.000000
S     0.000000   -1.054088    0.000000
C     1.270100    0.165512    0.000000
C     0.706100    1.441412    0.000000
H    -1.310300    2.352412    0.000000
H     1.310300    2.352412    0.000000
C     3.219600   -1.495488    0.000000
C     2.656500   -0.219888    0.000000
S     3.927900    0.998512    0.000000
C     5.196000   -0.220788    0.000000
C     4.632800   -1.495088    0.000000
H     2.615500   -2.406588    0.000000
H     5.236800   -2.406288    0.000000
C     7.149500    1.439512    0.000000
C     6.587100    0.165912    0.000000
S     7.858600   -1.051488    0.000000
C     9.102000    0.159612    0.000000
C     8.572000    1.432812    0.000000
H     6.546000    2.351012    0.000000
H    10.148900   -0.142688    0.000000
H     9.185200    2.337012    0.000000
          

\end{verbatim}


\end{document}